\PassOptionsToPackage{pdfauthor={Vladimir V. Kisil},%
    pdftitle={p-Mechanic and Field Theory},%
    pdfsubject={},%
    backref=page,%
  pdfkeywords={symmetries},
  hypertex
}{hyperref}
\documentclass[a4paper,reqno]{amsart}
        \newtheorem{thm}{Theorem}[section]
    
    \newtheorem{lem}[thm]{Lemma}

    \theoremstyle{definition}
    \newtheorem{defn}[thm]{Definition}

       \theoremstyle{remark}
       \newtheorem{rem}[thm]{Remark}

\providecommand{\href}[2]{#2}
\IfFileExists{eulervm.sty}{\usepackage{eulervm}
}{}
\newcommand{\eprint}[2]{E-print: \href{#1}{\texttt{#2}}}
\newcommand{\such}{\,\mid\,}
\providecommand{\comment}[1]{}
\providecommand{\myhbar}{h}
\newcommand{\Cliff}[2][\comment]{{\ensuremath{\mathcal{C}\kern-0.12em\ell(#1,#2)}}}
  \providecommand{\MR}[1]{\textbf{MR}~\href{http://www.ams.org/mathscinet-getitem?mr=#1}{\#~#1}}
  \newcommand{\Zbl}[1]{\textbf{Zbl}~\href{http://www.emis.de:80/cgi-bin/zmen/ZMATH/en/zmathf.html?first=1&maxdocs=3&type=html&an=#1&format=complete}{\#~#1}}

\newcommand{\scalar}[3][\relax]{\left\langle #2,#3 
        \right\rangle\ifx#1\relax\else_{#1}\fi}
\newcommand{\modulus}[2][\relax]{\left| #2 \right|\ifx#1\relax\else_{#1}\fi}
  \newcommand{\FSpace}[3][]{\ensuremath{\ifx#2l \ell_{#3}^{#1}{}\else
  #2_{#3}^{#1}{}\fi}} 
\newcommand{\Space}[3][]{\ensuremath{\mathbb{#2}^{#3}_{#1}{}}}
\newcommand{\algebra}[1]{\ensuremath{\mathfrak{#1}}}
\newcommand{\object}[2][\,]{\ensuremath{\mathrm{#2}#1}}
\providecommand{\fl}{}

\providecommand{\keywords}[1]{}
\providecommand{\urladdr}[1]{}
\providecommand{\ams}[1]{\subjclass[2000]{#1}}
\providecommand{\rmi}{\mathrm{i}}

\newcommand{\symp}[2]{#1\circ #2}
\providecommand{\myhbar}{h}
\providecommand{\orbit}[1]{\mathcal{O}_{#1}}
\providecommand{\uir}[1]{\rho_{#1}}

\providecommand{\e}[1]{\mathrm{e}^{#1}}

\newcommand{\anti}{\mathcal{A}}

\newcommand{\ub}[3][]{\left\{\!#1\left[#2,#3\right]\!#1\right\}}

\begin{document}

\title{$p$-Mechanics and Field Theory}
\author[Vladimir V. Kisil]%
{\href{http://maths.leeds.ac.uk/~kisilv/}{Vladimir
    V. Kisil}}
\thanks{On leave from the Odessa University.}
\email{kisilv@maths.leeds.ac.uk}
\address{%
School of Mathematics,
University of Leeds,
Leeds LS2\,9JT,
UK}


\urladdr{\href{http://maths.leeds.ac.uk/~kisilv/}%
{http://maths.leeds.ac.uk/\~{}kisilv/}}

\begin{abstract}
  The orbit method of Kirillov is used to derive  
  the \(p\)-mechanical brackets~\cite{Kisil00a}. 
  They generate the quantum
  (Moyal) and classical (Poisson) brackets on respective orbits corresponding to
  representations of the Heisenberg group. 
  The extension of \(p\)-mechanics to field theory is made through the
  De~Donder--Weyl Hamiltonian formulation. The principal step is the
  substitution of the Heisenberg group with Galilean. 
\end{abstract}
\ams{81R05, 81R15, 81T70, 
81S10, 81S30, 22E27, 22E70, 43A65}

\maketitle

\tableofcontents

\section{Introduction}
\label{sec:introduction}

The paper deals with quantization and bracket structures for
mechanical and field-theoretic problems. It extends certain mechanical
problems onto the field-theoretic framework, using the
finite-dimensional ``canonical formalism'' based on the concept of
\emph{polymomenta}~\cite{Kanatchikov95a,Kanatchikov98a,Kanatchikov98b,%
  Kanatchikov98c,Kanatchikov98d,Kanatchikov99a,Kanatchikov00a,%
  Kanatchikov01a,Kanatchikov01b}. This finite-dimensional framework is strongly
related to the multisymplectic formalism. In this way, the
infinite-dimensional symplectic framework and infinite-dimensional
canonical formalism traditionally used in field theory \emph{are
  avoided}. Our approach and results have some relationships with
above cited papers by I.V.~Kanatchikov.

On mechanical level the main idea is to use the Kirillov orbit method
in the case of Heisenberg
group~\cite{Kisil96a,Kisil00a,Kisil02e,Brodlie03a,BrodlieKisil03a}.
\(p\)-Mechanical brackets are constructed using the group algebra
(convolution algebra ) on the Heisenberg group. This brackets reduce
to the quantum (Moyal) or mechanical (Poisson) brackets, depending on
the used orbit.

In this paper generalized brackets are constructed for field
theory formulated in finite-dimensional polymomenta terms. The main
idea there is to use the Galilei group instead the Heisenberg group.
The constructed brackets also may be considered as some
field-theoretic Moyal-like brackets (involving several ``Planck
constants'') or classical brackets. This is similar to our construction
of \(p\)-mechanical brackets. The main difference is
that the Heisenberg group is replaced by the Galilei group and some
Clifford-algebra-valued quantities are used.

The paper outline is as follows. We start in
Section~\ref{sec:preliminaries} from the Heisenberg group and its
representations derived through the orbit method of Kirillov. In
Section~\ref{sec:conv-algebra-hg} we define \(p\)-mechanical
observables as convolutions on the Heisenberg group \(\Space{H}{n}\)
and study their commutators. We modify the commutator of two
\(p\)-observables by the antiderivative to the central vector field in
the Heisenberg Lie algebra in Section~\ref{sec:p-mechanical-bracket};
this produces \(p\)-mechanical brackets and corresponding dynamic
equation. Then the \(p\)-mechanical construction is extended to the
De~Donder--Weyl Hamiltonian formulation of the field
theory~\cite{Kanatchikov95a,Kanatchikov98a,Kanatchikov98b,%
Kanatchikov98c,Kanatchikov98d,Kanatchikov99a,Kanatchikov00a,%
Kanatchikov01a,Kanatchikov01b} in
Section~\ref{sec:de-donder-weyl-1}. To this end we replace the
Heisenberg group by the Galilean group in
Section~\ref{sec:segal-bargmann-type}. Expanded presentation of
Section~\ref{sec:introduction-into-p} could be found
in~\cite{Kisil02e}. Development of material from
Section~\ref{sec:de-donder-weyl} will follow in subsequent papers.

\section{Elements of $p$-Mechanics}
\label{sec:introduction-into-p}

\subsection{The Heisenberg Group and Its Representations}
\label{sec:preliminaries}

Let \((s,x,y)\), where \(x\), \(y\in \Space{R}{n}\) and \(s\in\Space{R}{}\), be
an element of the Heisenberg group
\(\Space{H}{n}\)~\cite{Folland89,Howe80b}. The group law on
\(\Space{H}{n}\) is given as follows:
\begin{equation}
  \label{eq:H-n-group-law}
  (s,x,y)*(s',x',y')=(s+s'+\frac{1}{2} \omega(x,y;x',y'),x+x',y+y'), 
\end{equation} 
where the non-commutativity is made by \(\omega\)---the
\emph{symplectic form}~\cite[\S~37]{Arnold91} on \(\Space{R}{2n}\): 
\begin{equation}
  \label{eq:symplectic-form}
  \omega(x,y;x',y')=xy'-x'y.
\end{equation}
The Lie algebra \(\algebra{h}^n\) of \(\Space{H}{n}\) is spanned by
left-invariant vector fields 
\begin{equation}
  S={\partial_s}, \qquad
  X_j=\partial_{ x_j}-{y_j}/{2}{\partial_s},  \qquad
 Y_j=\partial_{y_j}+{x_j}/{2}{\partial_s}
  \label{eq:h-lie-algebra}
\end{equation}
on \(\Space{H}{n}\) with the Heisenberg \emph{commutator relations} 
\begin{equation}
  \label{eq:heisenberg-comm}
[X_i,Y_j]=\delta_{i,j}S
\end{equation}
and  all other commutators vanishing. 
The exponential map \(\exp:
\algebra{h}^n\rightarrow \Space{H}{n}\) respecting
the multiplication~\eqref{eq:H-n-group-law} and
Heisenberg commutators is
\begin{displaymath}
  \exp: sS+\sum_{k=1}^n (x_kX_k+y_kY_k) \mapsto (s,x_1,\ldots,x_n,y_1,\ldots,y_n).
\end{displaymath} 

As any group \(\Space{H}{n}\) acts on itself by the conjugation automorphisms
\(\object{A}(g) h= g^{-1}hg\), which preserve the unit \(e\in
\Space{H}{n}\). The differential \(\object{Ad}: \algebra{h}^n\rightarrow
\algebra{h}^n\) of \(\object{A}\) at \(e\) is a linear map which could
be differentiated again to the representation \(\object{ad}\) of the Lie algebra
\(\algebra{h}^n\) by the
commutator: \(\object{ad}(A): B \mapsto [B,A]\). The dual space
\(\algebra{h}^*_n\) to the Lie algebra 
\(\algebra{h}^n\) is realised by the left invariant first order
differential forms on \(\Space{H}{n}\). By the duality
between \(\algebra{h}^n\) and \(\algebra{h}^*_n\) the map \(\object{Ad}\)
generates the \emph{co-adjoint
representation}~\cite[\S~15.1]{Kirillov76} \(\object{Ad}^*: \algebra{h}^*_n
\rightarrow \algebra{h}^*_n\):
\begin{equation}
  \label{eq:co-adjoint-rep}
  \object{ad}^*(s,x,y): (\myhbar ,q,p) \mapsto (\myhbar , q+\myhbar y,
  p-\myhbar x), \quad
  \textrm{where } (s,x,y)\in \Space{H}{n} 
\end{equation}
and \((\myhbar ,q,p)\in\algebra{h}^*_n\) in bi-orthonormal coordinates to
the exponential ones on \(\algebra{h}^n\).  There are two
types of orbits in~\eqref{eq:co-adjoint-rep} for \(\object{Ad}^*\)---Euclidean 
spaces \(\Space{R}{2n}\) and single points:
\begin{eqnarray}
  \label{eq:co-adjoint-orbits-inf}
  \orbit{\myhbar} & = & \{(\myhbar, q,p): \textrm{ for a fixed
  }\myhbar\neq 0 \textrm{ and  all } (q,p) \in  \Space{R}{2n}\}, \\
  \label{eq:co-adjoint-orbits-one}
  \orbit{(q,p)} & = & \{(0,q,p): \textrm{ for a fixed } (q,p)\in \Space{R}{2n}\}.
\end{eqnarray} The \emph{orbit method} of
Kirillov~\cite[\S~15]{Kirillov76} starts from the observation that the
above orbits parametrise irreducible unitary representations of
\(\Space{H}{n}\). All representations are
\emph{induced}~\cite[\S~13]{Kirillov76} by a character
\(\chi_\myhbar(s,0,0)=\e{2\pi \rmi \myhbar s}\) of the centre of
\(\Space{H}{n}\) generated by \((\myhbar,0,0)\in\algebra{h}^*_n\) and
shifts~\eqref{eq:co-adjoint-rep} from the \emph{left} on
orbits. Using~\cite[\S~13.2, Prob.~5]{Kirillov76} we get a neat
formula, which (unlike many other in literature) respects all
\emph{physical units}~\cite{Kisil02e}: 
\begin{equation}
  \textstyle
  \label{eq:stone-inf}
  \uir{\myhbar}(s,x,y): f_\myhbar (q,p) \mapsto 
  \e{ -2\pi\rmi( \myhbar s+qx+py)}
  f_\myhbar (q-\frac{\myhbar}{2} y, p+\frac{\myhbar}{2} x).
\end{equation}
The derived representation \(d\uir{\myhbar}\) of  the Lie algebra
\(\algebra{h}^n\) defined on the vector
fields~\eqref{eq:h-lie-algebra} is:
\begin{equation} 
  \textstyle 
  \fl
  d\uir{\myhbar}(S)=-2\pi\rmi  \myhbar I, \quad 
  d\uir{\myhbar}(X_j)=\myhbar \partial_{p_j}+ \frac{\rmi}{2}  q_j I ,\qquad
  d\uir{\myhbar}(Y_j)=-\myhbar \partial_{q_j}+ \frac{\rmi}{2}  p_j I
  \label{eq:der-repr-h-bar}
\end{equation}
Operators \(D^j_\myhbar\),  \(1\leq j\leq n\) representing vectors 
from the complexification of \(\algebra{h}^n\):
\begin{equation}
  \label{eq:Cauchy-Riemann}
  \textstyle \fl
  D^j_\myhbar =d\uir{\myhbar}(-X_j+ \rmi  Y_j)
  =\frac{\myhbar}{2}  (\partial_{p_j}+ \rmi \partial_{q_j})+2\pi(p_j+ \rmi  q_j) I
  ={\myhbar} \partial_{\bar{z}_j}+2\pi z_j I 
\end{equation} where \( z_j =p_j+ \rmi  q_j\) are used to give the
following classic result in terms of orbits:
\begin{thm}[Stone--von Neumann,
  cf. \textup{\cite[\S~18.4]{Kirillov76}, \cite[Chap.~1, \S~5]{Folland89}}] 
  \label{th:Stone-von-Neumann} 
  All unitary irreducible representations of \(\Space{H}{n}\) are
  parametrised up to equivalence by two classes of
  orbits~\eqref{eq:co-adjoint-orbits-inf}
  and~\eqref{eq:co-adjoint-orbits-one} of the co-adjoint
  representation~\eqref{eq:co-adjoint-rep} in \(\algebra{h}^*_n\): 
  \begin{enumerate}
  \item The infinite dimensional representations by transformation
    \(\uir{\myhbar}\)~\eqref{eq:stone-inf} for \(\myhbar \neq 0\) in
    Fock~\textup{\cite{Folland89,Howe80b}} space
    \(\FSpace{F}{2}(\orbit{\myhbar})\subset\FSpace{L}{2}(\orbit{\myhbar})\)
    of null solutions to the operators \(D^j_\myhbar\)
    \eqref{eq:Cauchy-Riemann}:
    \begin{equation}
      \label{eq:Fock-type-space}
      \FSpace{F}{2}(\orbit{\myhbar})=\{f_{\myhbar}(p,q) \in
      \FSpace{L}{2}(\orbit{\myhbar}) \such D^j_\myhbar f_{\myhbar}=0,\
       1 \leq j \leq n\}.
    \end{equation}
  \item The one-dimensional representations as multiplication by
    a constant on \(\Space{C}{}=\FSpace{L}{2}(\orbit{(q,p)})\) which
    drops out from~\eqref{eq:stone-inf} for \(\myhbar =0\):
    \begin{equation}
      \label{eq:stone-one}
      \uir{(q,p)}(s,x,y): c \mapsto \e{-2\pi \rmi(qx+py)}c.
    \end{equation}
  \end{enumerate}
\end{thm} 
Note that \(f_\myhbar(p,q)\) is in
\(\FSpace{F}{2}(\orbit{\myhbar})\) if and only if the function
\(f_\myhbar(z)\e{-\modulus{z}^2/\myhbar}\), \(z=p+\rmi q\) is in the
classical Segal--Bargmann space~\cite{Folland89,Howe80b}, particularly
is analytical in \(z\). Furthermore the space
\(\FSpace{F}{2}(\orbit{\myhbar})\) is spanned by the Gaussian
\emph{vacuum vector} \(\e{-2\pi(q^2+p^2)/\myhbar}\) and all
\emph{coherent states}, which are ``shifts'' of the vacuum vector by
operators~\eqref{eq:stone-inf}.

Commutative representations~\eqref{eq:stone-one} correspond to the
case \(\myhbar=0\) in the formula~\eqref{eq:stone-inf}. They are always neglected,
however their union naturally (see the appearance of Poisson
brackets in~\eqref{eq:Poisson}) acts as the classical \emph{phase space}:
\begin{equation}
  \label{eq:orbit-0}
  \orbit{0}=\bigcup_{(q,p)\in\Space{R}{2n}} \orbit{(q,p)}.
\end{equation}
Furthermore the structure of orbits of \(\algebra{h}_n^*\)
is implicitly present in equation~\eqref{eq:p-equation} and its symplectic
invariance~\cite{Kisil02e}.  

\subsection{Convolution and Commutator on $\Space{H}{n}$}
\label{sec:conv-algebra-hg}

Using  a left invariant measure \(dg\) on \(\Space{H}{n}\) the linear space
\(\FSpace{L}{1}(\Space{H}{n},dg)\)  can be upgraded 
to an algebra with the convolution multiplication:
\begin{equation}
  \fl
  (k_1 * k_2) (g) = \int_{\Space{H}{n}} k_1(g_1)\, k_2(g_1^{-1}g)\,dg_1 =
  \int_{\Space{H}{n}} k_1(gg_1^{-1})\, k_2(g_1)\,dg_1.
  \label{eq:de-convolution}
\end{equation}
Inner \emph{derivations} \(D_k\), \(k\in\FSpace{L}{1}(\Space{H}{n})\)
of \(\FSpace{L}{1}(\Space{H}{n})\) are given 
by the \emph{commutator} for \(f\in\FSpace{L}{1}(\Space{H}{n})\):
\begin{equation}
  \fl
  D_k: f \mapsto [k,f]=k*f-f*k
  =\int_{\Space{H}{n}} k(g_1)\left(
    f(g_1^{-1}g)-f(gg_1^{-1})\right)\,dg_1.
  \label{eq:commutator}
\end{equation}
A unitary representation \(\rho_\myhbar \) of \(\Space{H}{n}\) extends
 to \(\FSpace{L}{1}(\Space{H}{n} ,dg)\) by the formula:
\begin{eqnarray}
  \fl
\lefteqn{  [\rho_\myhbar (k)f](q,p)
 = \int_{\Space{H}{n}} k(g)\rho_\myhbar
  (g)f(q,p)\,dg \nonumber} \\
  &=&\int_{\Space{R}{2n}} \left(\int_{\Space{R}{}} k(s,x,y)\e{-2\pi\rmi \myhbar
      s}\,ds \right)
  \e{ -2\pi{\rmi}(qx+py)}      f (q-\myhbar y, p+\myhbar x) \,dx\,dy,
  \label{eq:rho-extended-to-L1}
\end{eqnarray} thus \(\rho_\myhbar (k)\) for a fixed \(\myhbar \neq
0\) depends only on \(\hat{k}_s(\myhbar,x,y)\)---the partial Fourier
transform \(s\rightarrow \myhbar\) of \(k(s,x,y)\). Then the
representation of the composition of two convolutions depends only
from
\begin{eqnarray}
  \lefteqn{
  (k'*k)\hat{_s}(\myhbar,x,y) 
  = 
  \int_{\Space{R}{}}\int_{\Space{H}{n}}k'(s',x',y')} \nonumber \\
 &&\qquad \times k(s-s'-\frac{1}{2}
 (xy'-yx'),x-x',y-y')\,ds'dx'dy'\e{ 2\pi\rmi \myhbar s}\, ds\nonumber \\
  &=& 
  \int_{\Space{R}{}}\int_{\Space{H}{n}}k'(s',x',y') \e{ 2\pi\rmi \myhbar
    s'} \nonumber \\
 &&\qquad \times k(s-s'-\frac{1}{2}
 (xy'-yx'),x-x',y-y')\e{  2\pi\rmi \myhbar (s-s')}\,ds'dx'dy' ds \nonumber \\
  &=& 
  \int_{\Space{R}{2n}}   \hat{k}'_s(\myhbar ,x',y')\,
  \hat{k}_s(\myhbar ,x-x',y-y')\,\e{ \pi{\rmi \myhbar}
    (xy'-yx')}\,dx'dy'. 
  \label{eq:commut-part-transform}
\end{eqnarray}
The last expression for the full Fourier transforms of \(k'\) and
\(k\) turn out to be the \emph{star product} known in {deformation
  quantisation}, cf. \cite[(9)--(13)]{Zachos02a}.  Consequently the
representation of the commutator~\eqref{eq:commutator} depends only
from~\cite{Kisil02e}:
\begin{eqnarray}
    \lefteqn{[k',k]\hat{_s}
    =
  \int_{\Space{R}{2n}}\hat{k}'_s(\myhbar ,x',y')
  \hat{k}_s(\myhbar ,x-x',y-y')}
 \nonumber  \\
  && \qquad \qquad \qquad \times 
  \left(\e{ \pi\rmi \myhbar  (xy'-yx')}-\e{-\pi\rmi \myhbar  (xy'-yx')}\right)\,dx'dy'
 \nonumber \\
 &=&   2 \rmi \! \int_{\Space{R}{2n}} 
 \hat{k}'_s(\myhbar ,x',y')
 \hat{k}_s(\myhbar ,x-x',y-y')\,\sin({\pi\myhbar}  (xy'-yx'))\,dx'dy',
\label{eq:repres-commutator}
\end{eqnarray}
which turn out to be exactly the ``Moyal brackets''~\cite{Zachos02a}
for the full Fourier transforms of \(k'\) and \(k\). Also the
expression~\eqref{eq:repres-commutator} vanishes for \(\myhbar=0\) as
can be expected from the commutativity of
representations~\eqref{eq:stone-one}.

\subsection{$p$-Mechanical Brackets on $\Space{H}{n}$}
\label{sec:p-mechanical-bracket}

An anti-derivation operator \(\anti\), which is a scalar multiple of a
right inverse of the vector field \(S\)~\eqref{eq:h-lie-algebra} on
\(\Space{H}{n}\), is defined by:
\begin{equation}
  S\anti=4\pi^2 I, \qquad \textrm{ where }\quad
  \label{eq:def-anti}
  \anti \e{ 2\pi\rmi \myhbar s}=\left\{ 
    \begin{array}{ll} \displaystyle
      \frac{2\pi}{\rmi\myhbar\strut} \strut \e{2\pi\rmi\myhbar s}, & \textrm{if } \myhbar\neq 0,\\
      4\pi^{2\strut} s, & \textrm{if } \myhbar=0.
    \end{array}
    \right. 
\end{equation} 
It can be extended by the linearity to
\(\FSpace{L}{1}(\Space{H}{n})\). We introduce~\cite{Kisil00a} a modified
convolution operation \(\star\) on \(\FSpace{L}{1}(\Space{H}{n})\):
\begin{displaymath}
    k_1\star k_2= k_1*(\anti k_2), 
\end{displaymath}
 and the associated modified commutator
 \begin{equation}
   \label{eq:star-and-brackets}
    \ub{k_1}{k_2}=k_1\star    k_2-k_2\star k_1,
 \end{equation}
 which will be called \emph{\(p\)-mechanical brackets} for reasons
 explained in~\eqref{eq:p-equation}.

One gets from~\eqref{eq:rho-extended-to-L1} the representation
\(\uir{\myhbar}(\anti k)=(i\myhbar)^{-1}\uir{\myhbar}(k)\) for any
\(\myhbar\neq 0\). Consequently the modification
of~\eqref{eq:repres-commutator} for \(\myhbar\neq0\) is only slightly
different from the original one:
\begin{equation}
  \fl
    \ub{k'}{k}\!\hat{_s}
    =   \int_{\Space{R}{2n}}
    \frac{2\pi}{\myhbar}\sin(\pi\myhbar  (xy'-yx'))\,
    \hat{k}'_s(\myhbar ,x',y')\,
    \hat{k}_s(\myhbar ,x-x',y-y') \,dx'dy',
\label{eq:repres-ubracket}
\end{equation}
However the last expression for \(\myhbar=0\) is significantly distinct
from the vanishing~\eqref{eq:repres-commutator}. From the
natural assignment \(\frac{4\pi}{\myhbar}\sin(\pi\myhbar (xy'-yx'))=4\pi^2(xy'-yx')\) for
\(\myhbar=0\) we get the Poisson brackets for the Fourier
transforms of \(k'\) and \(k\) defined on \(\orbit{0}\)~\eqref{eq:orbit-0}:
\begin{equation}
  \label{eq:Poisson}
    \uir{(q,p)}\ub{k'}{k} = \frac{\partial \hat{k}'}{\partial q}
    \frac{\partial \hat{k}}{\partial p}
    -\frac{\partial \hat{k}'}{\partial p} \frac{\partial \hat{k}}{\partial q}.
\end{equation}
Furthermore the dynamical equation based on the modified
commutator~\eqref{eq:star-and-brackets} with a suitable Hamilton type
function \(H(s,x,y)\) for an observable \(f(s,x,y)\) on
\(\Space{H}{n}\) 
\begin{equation}
  \fl
  \label{eq:p-equation}
  \dot{f}=\ub{H}{f} \textrm{ is reduced } \left\{ 
    \begin{array}{l}
      \mbox{by \(\uir{\myhbar}\), \(\myhbar\neq0\) on
        \(\orbit{\myhbar}\)~\eqref{eq:co-adjoint-orbits-inf} to Moyal's
        eq. \cite[(8)]{Zachos02a};}\\
      \mbox{by \(\uir{(q,p)}\) on
        \(\displaystyle\orbit{0}\)~\eqref{eq:orbit-0} to Poisson's 
        eq. \cite[\S~39]{Arnold91}.}
    \end{array}
  \right.
\end{equation}
The same relationships are true for the solutions of these three equations,
see~\cite{Kisil00a} for the harmonic oscillator
and~\cite{Brodlie03a,BrodlieKisil03a} for forced oscillator examples.

\section{De~Donder--Weyl Field Theory}
\label{sec:de-donder-weyl}

We extend \(p\)-mechanics to the De~Donder--Weyl field theory,
see~\cite{Kanatchikov95a,Kanatchikov98a,Kanatchikov98b,%
Kanatchikov98c,Kanatchikov98d,Kanatchikov99a,Kanatchikov00a,%
Kanatchikov01a,Kanatchikov01b}
for detailed exposition and further references. We will be limited
here to the preliminary discussion which extends the comment 5.2.(1)
from the earlier paper~\cite{Kisil02e}. Our notations will slightly
different from the used in the
papers~\cite{Kanatchikov95a,Kanatchikov98a,Kanatchikov98b,%
Kanatchikov98c,Kanatchikov98d,Kanatchikov99a,Kanatchikov00a,%
Kanatchikov01a,Kanatchikov01b}
to make it consistent with the used above and avoid clashes.

\subsection{Hamiltonian Form of Field Equation}
\label{sec:de-donder-weyl-1}

Let the underlying space-time has the dimension \(n+1\) and be parametrised
by coordinates \(u^\mu\), \(\mu=0,1,\ldots,n\) (with \(u^0\) parameter
traditionally associated with a time-like direction). Let us consider a field 
described by some \(m\)-component tensor \(q^a\), \(a=1,\ldots,m\). For a
system defined by a Lagrangian density \(L(q^a, \partial_\mu q^a,
u^\mu)\) De~Donder--Weyl theory suggests new set of \emph{polymomenta}
\(p_a^\mu\) and \emph{DW Hamiltonian function}
\(H(q^a,p_a^\mu,u^\mu)\) defined as follows:
\begin{equation} \label{eq:polymomenta}
  p_a^\mu:=\frac{\partial L (q^a, \partial_\mu q^a,
u^\mu)}{\partial  (\partial_\mu y^a)}\quad
  \textrm{ and } \quad
  H(q^a,p_a^\mu,u^\mu)=p_a^\mu\,\partial_\mu q^a-L(q^a, \partial_\mu
  q^a, u^\mu). 
\end{equation}
A multidimensional variational problem for the Lagrangian \(L(q^a, \partial_\mu q^a,
u^\mu)\) leads to the  Euler--Lagrange field equations:
\begin{displaymath}
  \frac{d}{d u^{\mu}}\left( \frac{\partial L}{\partial (\partial_\mu
      q^a)}\right) - \frac{\partial L}{\partial q^a}=0.
\end{displaymath}
Just as in particle mechanics polymomenta~\eqref{eq:polymomenta}
help us to transform the above Euler--Lagrange field equations 
to the Hamilton form:
\begin{equation}
  \label{eq:hamilton-field}
  \frac{\partial q^a}{\partial u^\mu}=\frac{\partial H}{\partial
    p_a^\mu}, \qquad
  \frac{\partial p_a^\mu}{\partial u^\mu}=-\frac{\partial H}{\partial
    q^a},
\end{equation} with the standard summation (over repeating
\emph{Greek} indexes)
convention. The main distinction from a particle mechanics is the
existence of \(n+1\) different polymomenta \(p_a^\mu\) associated to
each field variable \(q^a\). Therefore, the particle mechanics could
be considered as a particular case when \(n+1\) dimensional space-time
degenerates for \(n=0\) to ``time only''.

The next two natural
steps~\cite{Kanatchikov95a,Kanatchikov98a,Kanatchikov98b,%
Kanatchikov98c,Kanatchikov98d,Kanatchikov99a,Kanatchikov00a,%
Kanatchikov01a,Kanatchikov01b}
inspired by particle mechanics are:
\begin{enumerate}
\item Introduce an appropriate Poisson structure, such that the Hamilton
  equations~\eqref{eq:hamilton-field} will represent the Poisson
  brackets.
\item Quantise the above Poisson structure by some means,
  e.g. Dirac-Heisenberg-Shr\"odinger-Weyl technique or geometric
  quantisation.   
\end{enumerate}

We use here another path: first to construct a \(p\)-mechanical model
for equations~\eqref{eq:hamilton-field} and then deduce its quantum
and classical derivatives as was done for the particle mechanics
above. To simplify presentation we will start from the scalar field,
i.e. \(m=1\). Thus we drop index \(a\) in \(q^a\) and \(p_a^\mu\) and
simply write \(q\) and \(p^\mu\) instead.  

We also assume that the underlying space-time is flat with a constant
metric tensor \(\eta^{\mu\nu}\). This metric defines a related
\emph{Clifford algebra}~\cite{BraDelSom82,Cnops02a,DelSomSou92} with
generators \(e_\mu\) satisfying the relations
\begin{equation}
  \label{eq:clifford-def}
  e_\mu e_\nu + e_\nu e_\mu = \eta^{\mu\nu}.
\end{equation}
\begin{rem}
  For the Minkowski space-time (i.e. in the context of special
  relativity) a preferable choice may be
  \emph{quaternions}~\cite{Sudbery79} with generators \(i\), \(j\),
  \(k\) instead the general Clifford algebra.
\end{rem}
\begin{rem}
  To avoid the possibility of confusion with imaginary units \(e_\nu\) we
  will use the another font for the base of natural logarithms
  \(\e{}\).
\end{rem}

Since \(q\) and \(p^\mu\) look like conjugated variables
\(p\)-mechanics suggests that they should generate a Lie algebra with
relations similar to~\eqref{eq:heisenberg-comm}. The first natural
assumption is the \(n+3(=1+(n+1)+1)\)-dimensional Lie algebra spanned
by \(X\), \(Y_\mu\), and \(S\) with the only non-trivial commutators
\([X,Y_\mu]=S\). However as follows from the Kirillov
theory~\cite{Kirillov62} any its unitary irreducible representation is
limited to a representation of \(\Space{H}{1}\) listed by the
Stone--von Neumann Theorem~\ref{th:Stone-von-Neumann}. Consequently
there is a little chance that we could obtain the field
equations~\eqref{eq:hamilton-field} in this way.

\subsection{Convolutions and  Commutator on $\Space{G}{n+1}$}
\label{sec:segal-bargmann-type}

The next natural candidate is the Galilean group \(\Space{G}{n+1}\),
i.e. a nilpotent step 2 Lie group with
the \(2n+3(=1+(n+1)+(n+1))\)-dimensional
Lie algebra. It has a basis \(X\), \(Y_\mu\), and \(S_\mu\)
with \(n+1\)-dimensional centre spanned by \(S_\mu\). The only
non-trivial commutators  are
\begin{equation}
  \label{eq:galelean-comm}
  [X,Y_\mu]=S_\mu, \qquad \textrm{ where } \mu=0,1,\ldots,n.
\end{equation} Again the Kirillov theory~\cite{Kirillov62} assures
that any its \emph{complex-valued} irreducible representation is a
representation of \(\Space{H}{1}\). However the multidimensionality of
the centre of \(\Space{G}{n+1}\) offers an
option~\cite{CnopsKisil97a,Kisil01d} to consider \emph{Clifford
valued} representations of \(\Space{G}{n+1}\). 

\begin{rem}
  The appearance of Clifford algebra in connection with field theory
  and space-time geometry is natural. For example, the conformal
  invariance of space-time has profound consequences in
  astrophysics~\cite{Segal76} and, in their turn, conformal (M\"obius)
  transformations are most naturally represented by linear-fractional
  transformations in Clifford algebras~\cite{Cnops02a}. Some other
  links between nilpotent Lie groups and Clifford algebras are listed
  in~\cite{Kisil01d}.
\end{rem}

The Lie group \(\Space{G}{n+1}\) is a manifold homeomorphic to
\(\Space{R}{2n+3}\) with coordinates \((s,x,y)\), where
\(x\in\Space{R}{}\) and \(s,y\in\Space{R}{n+1}\). The group
multiplication in these coordinates is defined by,
cf.~\eqref{eq:H-n-group-law}:
\begin{eqnarray}
  \label{eq:Galilean-group-law}
  \lefteqn{(s,x,y)*(s',x',y')}\\
&=& \textstyle (s_0+s_0'+\frac{1}{2}
  \omega(x,y_0;x',y_0'),\ldots, s_n+s_n'+\frac{1}{2}
  \omega(x,y_n;x',y_n'),x+x',y+y'). \nonumber 
\end{eqnarray} 
The Lie algebra \(\algebra{g}^{n+1}\) is realised by the left (right)
invariant vector fields on \(\Space{G}{n+1}\) which satisfy
to~\eqref{eq:galelean-comm}: 
\begin{equation}
  \label{eq:Gn-v-f}
  S_{l(r)}^j=\pm\frac{\partial}{\partial s_j}, \quad
  X_{l(r)}=\pm\frac{\partial}{\partial x}-\sum_{j=0}^n
  \frac{y_j}{2}\frac{\partial}{\partial s_j}, \quad 
  Y_{l(r)}^j=\pm\frac{\partial}{\partial y_j}+
  \frac{x}{2}\frac{\partial}{\partial s_j}. 
\end{equation}

The dual \(\algebra{g}^*_{n+1}\) of the Lie algebra
\(\algebra{g}^{n+1}\) coincides with \(\Space{R}{2n+3}\) with
coordinates \((\myhbar_0,\ldots,\myhbar_n,q,p_0,\ldots,p_n)\) in the
basis biorthogonal to the basis \(S^j\), \(X\), \(Y^j\),
\(j=0,1,\ldots,n\) of \(\algebra{g}^{n+1}\). Similarly to the
realisation~\eqref{eq:Fock-type-space} of Fock type spaces on
\(\algebra{h}^{n}\) we define the vacuum vector
\(v_{{h},0}\) parametrised by the
\((n+1)\)-tuple of the Planck constants
\({h}=(h_0,h_1,\ldots,h_n)\) as follows (cf.~\cite[(2.25)]{Kisil02e}): 
\begin{equation}
  v_{{h},0}(s,x,y) = \sum_{j=0}^n 
  \exp 2\pi h_k \left({-e_j s_j
      -\textstyle
      \frac{1}{4}(x^2+y_0^2+y_1^2+\cdots +y_n^2)}\right). \label{eq:Gn-vacuum}
\end{equation} Here one can use the Euler formula
\(\e{a+e_jb}=\e{a}(\cos b + e_j\sin b)\) for \(a\), \(b\in\Space{R}{}\) as a definition.
The coherent states \(v_{h,g}\)
are obtained as left shifts of \(v_{h,0}\) by \(g=(s,x,y)\) on
\(\Space{G}{n+1}\):
\begin{eqnarray}
  v_{h,g}(g')&=&\uir{l}(s,x,y)v_{h,0}(s',x',y') \label{eq:Gn-coherent-states} \\
    &=& \sum_{j=0}^n \exp\,  2\pi h_j\!\left(-e_j\left(s_j'-s_j - 
        {
          \frac{1}{2}}
        \omega(x,y_j;x',y_j')\right)\right. \nonumber \\
      &&\qquad\qquad     \left. -
      \frac{1}{4}\left((x'-x)^2+\sum_{l=0}^n  (y'_l-y_l)^2\right)\right).
    \nonumber 
\end{eqnarray} 
Note that the vacuum vector~\eqref{eq:Gn-vacuum} and
coherent states~\eqref{eq:Gn-coherent-states} are different from ones
used in~\cite{CnopsKisil97a}.
\begin{defn}
  For a fixed parameter \(h=(h_0,h_1,\ldots,h_n)\in\Space{R}{n+1}\)
  we define Segal--Bargmann space \(\FSpace[h]{F}{2}(\Space{G}{n+1})\)
  of functions on \(\Space{G}{n+1}\) as  the closure of the
  linear span of vectors \(v_{h,g}\) for all \(g\in\Space{G}{n+1}\)
  in the norm defined by the inner product:
  \begin{eqnarray}
    \scalar[h]{f_1}{f_2}&=&\int_{\Space{R}{n+1}}\int_{\Space{R}{}}
    \sum_{j=0}^n 
    h_j\left(\int_{\Space{R}{n+1}}
    \bar{f}_1(s,x,y)\, \e{2\pi e_j h_j s_j}\,ds\right)\label{eq:Gn-scalar-prod}\\
  &&\qquad\qquad\qquad\times\left(
  \int_{\Space{R}{n+1}}\e{-2\pi e_j h_j s_j}\,{f}_2(s,x,y)\,ds\right)\, dx\,dy. \nonumber 
  \end{eqnarray} The linear combinations of coherent
  states~\eqref{eq:Gn-coherent-states} with Clifford coefficients
  multiplied from the \emph{right}.
\end{defn}
The following could be directly verified:
\begin{lem}
  \begin{enumerate}
  \item The left regular action of \(\Space{G}{n+1}\) is unitary with
    respect to the inner product~\eqref{eq:Gn-scalar-prod}.
  \item The vacuum vector~\eqref{eq:Gn-vacuum} and consequently all
    coherent states~\eqref{eq:Gn-coherent-states} have the unit norm
    in \(\FSpace[h]{F}{2}(\Space{G}{n})\).
  \end{enumerate}
\end{lem}

Let \(P_j\) be the operator
\(\FSpace[h]{F}{2}(\Space{G}{n})\rightarrow
\FSpace[h]{F}{2}(\Space{G}{n})\) defined through the partial
Fourier transform \(s_j\rightarrow h_j\) as follows:
\begin{displaymath}
  [P_j k](s,x,y)=e^{2\pi e_j s_j h_j}\int_{\Space{R}{n+1}} k(s,x,y)
  \,\e{-2\pi e_j s_j h_j}\,ds. 
\end{displaymath}
The following properties can be directly verified.
\begin{lem}
  If all components \(h_j\) of
  \(h=(h_0,h_1,\ldots,h_n)\in\Space{R}{n+1}\) are non-zero then
  \begin{enumerate}
  \item Each operator \(P_j\) is an orthogonal projection on
    \(\FSpace[h]{F}{2}(\Space{G}{n})\). 
  \item\label{item:P-resolution} The sum of all \(P_j\) is the identity operator on
    \(\FSpace[h]{F}{2}(\Space{G}{n})\): \(\sum_0^n P_j =I\). 
  \item All operators \(P_j\) commute with the left and right action
    of \(\Space{G}{n+1}\) on \(\FSpace[h]{F}{2}(\Space{G}{n})\).
  \end{enumerate}
\end{lem}

\begin{defn}
  Observables are defined as convolution \(K\) operators on
  \(\FSpace{L}{2}(\Space{G}{n+1})\) with a kernel \(k(s,x,y)\), which
  has a property:
  \begin{equation}\label{eq:kernel-definition}
    \int_{\Space{R}{n+1}} k(s,x,y)\, \e{-2\pi e_j h_j s_j}\, ds =
      \int_{\Space{R}{n+1}} k(s,x,y)\, \e{2\pi e_j h_j s_j}\, ds.
  \end{equation}
\end{defn} 
Due to the resolution \(\sum_0^n P_j =I\) on
\(\FSpace[h]{F}{2}(\Space{G}{n})\) from~\ref{item:P-resolution} we see that
the restriction of a convolution operator \(K = K\sum_0^n P_j
=\sum_0^n KP_j\) to \(\FSpace[h]{F}{2}(\Space{G}{n})\) is completely
defined by the set of operators \(KP_j\).  Kernels of operators
\(KP_j\) are given by partial
Fourier transforms \((s_0,\ldots,s_j,\ldots,s_n)\rightarrow
(0,\ldots,h_j,\ldots,0)\): 
\begin{displaymath}
  \hat{k}_j(0,\ldots,h_j,\ldots,0,x,y)=\int_{\Space{R}{n+1}}
  k(s_0,\ldots,s_n,x,y)\, \e{-2\pi e_j h_j s_j}\,ds_0\ldots ds_j \ldots ds_n.
\end{displaymath}

The following Lemma simplify further calculations:
\begin{lem}
  Let a kernel \(k(s,x,y)\) satisfy to the
  identity~\eqref{eq:kernel-definition}, then for any constant
  \(c\in\Cliff{n+1}\):
  \begin{displaymath}
    \int_{\Space{R}{n+1}} k(s,x,y)\, c\, \e{-2\pi e_j h_j s_j}\, ds =
      \int_{\Space{R}{n+1}} k(s,x,y)\, \e{-2\pi e_j h_j s_j}\, ds\, c.
  \end{displaymath}
\end{lem}
\begin{proof}
  In a trivial way we can present \(c\, \e{-2\pi e_j h_j s_j} =
  \e{-2\pi e_j h_j s_j}c_1 +\e{2\pi e_j h_j s_j} c_2\), where
  \(c_1=\frac{1}{2}(c-e_j c e_j)\) and \(c_2=\frac{1}{2}(c+e_j c
  e_j)\), thus \(c=c_1+c_2\). Then
  \begin{eqnarray}
  \lefteqn{      \int_{\Space{R}{n+1}} k(s,x,y)\, c\, \e{-2\pi e_j h_j
      s_j}\,ds} \nonumber \\
  &=&   \int_{\Space{R}{n+1}} k(s,x,y)\, \e{-2\pi e_j h_j
    s_j}\,ds\,c_1+ 
  \int_{\Space{R}{n+1}} k(s,x,y)\,\e{2\pi e_j h_j s_j}\,ds\,c_2
  \nonumber \\
  &=&   \int_{\Space{R}{n+1}} k(s,x,y)\, \e{-2\pi e_j h_j
    s_j}\,ds\,c_1+ 
  \int_{\Space{R}{n+1}} k(s,x,y)\,\e{-2\pi e_j h_j s_j}\,ds\,c_2 
  \label{eq:sgn-trans}\\
  &=&    \int_{\Space{R}{n+1}} k(s,x,y)\, \e{-2\pi e_j h_j s_j}\, ds\,
  c, \nonumber 
  \end{eqnarray}
  where transformation \eqref{eq:sgn-trans} follows from the
  property~\eqref{eq:kernel-definition}. 
\end{proof}

Using this Lemma we found that the restriction of the
composition \(k'* k\) of two convolutions to 
\(\FSpace[h]{F}{2}(\Space{G}{n})\) depends only
(cf.~\eqref{eq:commut-part-transform}) on:
\begin{eqnarray}
  (k'* k)\hat{_j} &=& \int_{\Space{R}{n+2}}
  \hat{k}'_j(0,\ldots,h_j,\ldots,0,x',y')\, \hat{k}_j\left(0, \ldots, h_j,
  \ldots, 0,x-x',y-y'\strut\right)\nonumber \\
  &&\qquad\times 
  \e{\pi e_j h_j(xy'_j-y_jx')} \,dx'\,dy'.\nonumber 
\end{eqnarray}

Thus the restriction of the commutator \(k'* k-k*k'\) of two
convolutions to \(\FSpace[h]{F}{2}(\Space{G}{n})\) depends only
(cf.~\eqref{eq:repres-commutator}) on: 
\begin{eqnarray}
  \lefteqn{(k'* k-k*k')\hat{_j}}\nonumber \\
  &=& \int_{\Space{R}{n+2}}
  \hat{k}'_j(0,\ldots,h_j,\ldots,0,x',y')\, \hat{k}_j\left(0, \ldots, h_j,
  \ldots, 0,x-x',y-y'\strut\right)\nonumber \\
  &&\qquad\times 
  \left(\e{\pi e_j h_j(xy'_j-y_jx')} - \e{-\pi e_j
      h_j(xy'_j-y_jx')}\right)dx'\,dy'.\nonumber \\
  &=& \int_{\Space{R}{n+2}}
  \hat{k}'_j(0,\ldots,h_j,\ldots,0,x',y')\, \hat{k}_j\left(0, \ldots, h_j,
  \ldots, 0,x-x',y-y'\strut\right)\nonumber \\
  &&\qquad\times 
  2e_j \sin\left(\pi
    h_j(xy'_j-y_jx')\right)dx'\,dy'. \label{eq:commut-restrict} 
\end{eqnarray}

\subsection{$p$-Mechanical Brackets on $\Space{G}{n+1}$}
\label{sec:p-mech-brack}

To define an appropriate brackets of two observables \(k'\) and \(k\)
we will again modify the restriction \([k',k]\hat{}_j\) of their
commutator on \(\FSpace[h]{F}{2}(\Space{G}{n+1})\). To this end we use
(cf.~\eqref{eq:def-anti} ) antiderivative operators \(\anti_0\),
\(\anti_1\), \ldots, \(\anti_n\) which are multiples of right inverse
to the vector fields \(S^0\), \(S^1\), \ldots,
\(S^n\)~\eqref{eq:Gn-v-f}:
\begin{equation}
  S^j\anti_j=4\pi^2 P_j, \quad \textrm{ where }
  \label{eq:def-anti-Galilean}
  \anti_j\e{ 2\pi e_k \myhbar_k s_k}=\left\{ 
    \begin{array}{ll} \displaystyle
      \frac{2\pi}{e_j\myhbar_j\strut} 
      \strut \e{2\pi e_j\myhbar_j s_j}, & \textrm{if } \myhbar_j\neq 0,
      \textrm{ and } j=k;\\
      4\pi^{2\strut} s_j, & \textrm{if } \myhbar_j=0,  \textrm{ and } j=k; \\
      0, & \textrm{if } \myhbar_k\neq 0,  \textrm{ and }
      j\neq k. \\
    \end{array}
    \right. 
\end{equation} 
The definition of the brackets follows the ideas
of~\cite[\S~3.3]{CnopsKisil97a}: to each vector field \(S_j\)
should be associated a generator \(e_j\) of Clifford
algebra~\eqref{eq:clifford-def}.  
Thus our brackets are as follows, cf.~\eqref{eq:star-and-brackets}:
\begin{equation}
  \label{eq:field-brackets}
  \ub{B_1}{B_2}=B_1* B_2 \anti -
  B_2*B_1 \anti, \qquad \textrm{ where }  \anti=\sum_{j=0}^n e_j\anti_j.
\end{equation}
These brackets will be used in the right-hand side of the \(p\)-dynamical
equation. Its left-hand side should contain a replacement for the time
derivative. As was already mentioned
in~\cite{Kanatchikov95a,Kanatchikov98a,Kanatchikov98b,%
Kanatchikov98c,Kanatchikov98d,Kanatchikov99a,Kanatchikov00a,%
Kanatchikov01a,Kanatchikov01b}
the space-time play a r\^ole of multidimensional time in the
De~Donder--Weyl construction. Thus we replace time derivative by the
symmetric pairing \(\symp{D}{}\) with the \emph{Dirac
  operator}~\cite{BraDelSom82,Cnops02a,DelSomSou92} \(D=e_j
\partial_j\) as follows:
\begin{equation}
  \label{eq:dirac}
  \symp{D}{f}=-\frac{1}{2}\left(e_j \frac{\partial f}{\partial
      u^j}  + \frac{\partial f}{\partial u^j} e_j \right),
  \qquad \textrm{ where } D=e_j \partial_j. 
\end{equation}
Finally the \(p\)-mechanical dynamical equation, cf.~\eqref{eq:p-equation}:
\begin{equation}
  \label{eq:p-mech-field-eq}
  \symp{D}{f}=\ub{H}{f},
\end{equation}
is defined through the brackets~\eqref{eq:field-brackets} and the
Dirac operator~\eqref{eq:dirac}.

To ``verify'' the equation~\eqref{eq:p-mech-field-eq} we will find its
classical representation and compare it
with equations~\eqref{eq:hamilton-field}. Indeed,
combining~\eqref{eq:commut-restrict} and~\eqref{eq:def-anti-Galilean}
for \(h_j\neq 0\) we got:
\begin{eqnarray}
  \ub{k'}{k}\hat{_j} &=& \int_{\Space{R}{n+2}}
  \hat{k}'_j(0,\ldots,h_j,\ldots,0,x',y')\, \hat{k}_j\left(0, \ldots, h_j,
  \ldots, 0,x-x',y-y'\strut\right)\nonumber \\
  &&\qquad\times 
  4\pi^2e_j\frac{\sin\left(\pi    h_j(xy'_j-y_jx')\right)}{\pi
    h_j}dx'\,dy'. 
  \label{eq:ub-restrict} 
\end{eqnarray}
In the limit \(h_j\rightarrow 0\) this naturally becomes 
\begin{eqnarray}
  \ub{k'}{k}\hat{_j} &=& \int_{\Space{R}{n+2}}
  \hat{k}'_j(0,\ldots,0,x',y')\, \hat{k}_j\left(0,\ldots,
    0,x-x',y-y'\strut\right)\nonumber \\ 
  &&\qquad\times 
  4\pi^2e_j(xy'_j-y_jx')dx'\,dy'. 
  \label{eq:ub-class-limit} 
\end{eqnarray}
Then the representation \(\uir{(q,p)}\) of the \(\ub{k'}{k}\hat{_j}\) is
expressed through the complete Fourier transform \((s,x,y)\rightarrow
(0,q,p)\) as follows:
\begin{displaymath}
  \uir{(q,p)}\left( \ub{k'}{k}\hat{_j}\right) = \sum_{j=0}^n\left(\frac{\partial
    \hat{k}'(0,q,p)} {\partial q}
    \frac{\partial \hat{k}(0,q,p)}{\partial p^j}
    -\frac{\partial \hat{k}'(0,q,p)}{\partial p^j} \frac{\partial
      \hat{k}(0,q,p)}{\partial q}\right) e_j. 
\end{displaymath}

Finally from the decomposition~\ref{item:P-resolution} it follows that 
\(\ub{k'}{k}=\sum_{j=0}^n \ub{k'}{k}\hat{}_j\) and 
we obtain the classical representation of \(p\)-mechanical brackets,
cf.~\eqref{eq:Poisson}: 
\begin{equation}
  \label{eq:field-Poisson}
    \uir{(q,p)}\ub{k'}{k} = \sum_{j=0}^n\left(\frac{\partial \hat{k}'} {\partial q}
    \frac{\partial \hat{k}}{\partial p^j}
    -\frac{\partial \hat{k}'}{\partial p^j} \frac{\partial
      \hat{k}}{\partial q}\right)e_j. 
\end{equation}
Consequently the dynamics from the
equation~\eqref{eq:p-mech-field-eq} of field observable \(q\) with a
scalar-valued Hamiltonian \(H\) in the classical representation is
given by: 
\begin{equation}
  \label{eq:field-q-eq}
  \symp{D}{q}= \sum_{j=0}^n\left(\frac{\partial H} {\partial q}
    \frac{\partial}{\partial p^j}
    -\frac{\partial H}{\partial p^j} \frac{\partial
      }{\partial q}\right) e_j  q \qquad \Longleftrightarrow  \qquad 
    \sum_{j=0}^n\frac{\partial q}{\partial u^j}e_j =
    \sum_{j=0}^n\frac{\partial H}{\partial p^j} e_j.
\end{equation} After the separation of components in the last equation
with different generators \(e_j\) we get the first \(n+1\) equations
from the set~\eqref{eq:hamilton-field}.

To obtain the last equation for polymomenta~\eqref{eq:hamilton-field} we
again use the Clifford algebra generators to construct the
\emph{combined polymomenta} \(p=\sum_0^n e_k p^k\). For them:
\begin{eqnarray*}
  \symp{D}{p} &=& -\sum_{j=0}^n \frac{1}{2}\left(e_j \frac{\partial
      \sum_{0}^n e_k p^k}{\partial 
      u^j}  + \frac{\partial\sum_{0}^n  e_k p^k}{\partial u^j} e_j
  \right) = -\sum_{j=0}^n\frac{\partial p^j}{\partial u^j} e_j e_j
  = \sum_{j=0}^n\frac{\partial p^j}{\partial u^j} , \\
  \ub{H}{p} &=& \sum_{j=0}^n\left(\frac{\partial H} {\partial q}
    \frac{\partial\sum_{0}^n  e_k p^k}{\partial p^j}
    -\frac{\partial H}{\partial p^j}  \frac{\sum_{0}^n \partial
      e_k p^k}{\partial q}\right)e_j = \sum_{j=0}^n\frac{\partial H}
  {\partial q}e_j e_j  = C\frac{\partial H}{\partial q} ,
\end{eqnarray*} where \(C=\sum_{0}^ne_j e_j\), i.e. \(C=-2\) for the
Minkowski space-time.  Thus the
equation~\eqref{eq:p-mech-field-eq} for the combined polymomenta
\(p=\sum_k e_k p^k\) in the classical representation becomes:
\begin{equation}
  \label{eq:field-p-eq}
    \sum_{j=0}^n\frac{\partial p^j}{\partial u^j}  =
    C\frac{\partial H}{\partial q} , 
\end{equation}
i.e. coincides with the last equation in~\eqref{eq:hamilton-field} up
to a constant factor \(-C=-\sum_{0}^ne_j e_j\). If this constant is
non-zero the second equation in~\eqref{eq:hamilton-field} is
equivalent to the equation~\eqref{eq:field-p-eq} with the Hamilton
function \(H_C(q,-\frac{1}{C}p)=-\frac{1}{C}\partial_j q p^j -L(q,
\partial_j q , u)\).
Thus \(p\)-mechanical equation~\eqref{eq:p-mech-field-eq} passed
the test by its classical representation.

Consequently we may assume that the ``quantum'' representations
\(\uir{h}=\uir{(h_0,\ldots,h_n)}\) of \(\Space{G}{n+1}\) map the
\(p\)-mechanical bracket \(\ub{k'}{k}\)~\eqref{eq:field-brackets} and
corresponding dynamic equation~\eqref{eq:p-mech-field-eq} to the
equation of quantum fields. Such an image \([\cdot,\cdot]_q\) of
\(\ub{\cdot}{\cdot}\) is easily obtained from~\eqref{eq:ub-restrict}
and the decomposition~\ref{item:P-resolution}:
\begin{eqnarray}
  \label{eq:ub-quantum-field} 
  \lefteqn{[\uir{h}(k'),\uir{h}(k)]_q  = \uir{h}\ub{k'}{k}}  \\
  &=&\int_{\Space{R}{n+2}} \sum_{j=0}^n
  \hat{k}'_j(0,\ldots,h_j,\ldots,0,x',y')\, \hat{k}_j\left(0, \ldots, h_j,
  \ldots, 0,x-x',y-y'\strut\right)\nonumber \\
  &&\qquad\times  \frac{4\pi e_j}{h_j}
  \sin\left(\pi    h_j(xy'_j-y_jx')\right)dx'\,dy'. \nonumber 
\end{eqnarray}
This brackets are similar to the Moyal
brackets~\eqref{eq:repres-ubracket}. Consequently the dynamic
equation~\eqref{eq:p-mech-field-eq} with a Hamiltonian \(\uir{h}(H)\)
for an observable \(\uir{h}(k)\) becomes in the quantum 
representation 
\begin{equation}
  \label{eq:quantum-field-eq}
  \symp{D}{\uir{h}(k)}=[\uir{h}(H),\uir{h}(k)]_q,
\end{equation} in the notations~\eqref{eq:ub-quantum-field}. Therefore
the image~\eqref{eq:quantum-field-eq} of the
equation~\eqref{eq:p-mech-field-eq} could stand for a quantisation of
its classical images in~\eqref{eq:hamilton-field},
\eqref{eq:field-Poisson}. A further study of quantum images of the
equation~\eqref{eq:p-mech-field-eq} as well as extension to vector or
spinor fields should follow in subsequent papers.  In different spaces
of functions these quantisations corresponds to
Dirac-Heisenberg-Schr\"odinger-Weyl, geometric, deformational,
etc. quantisations of particle mechanics.

\begin{rem}
  To consider vector or spinor fields with components \(q_a\),
  \(a=1,\ldots,m\) it worths to introduce another Clifford algebra
  with generators \(c^a\) and consider a composite field
  \(q=c^aq_a\). There are different ways to link Clifford and Grassmann
  algebras, see e.g.~\cite{Berezin86,Hestenes99}. Through such a link
  the Clifford algebra with generators \(e_j\) corresponds to
  horizontal differential forms in the sense
  of~\cite{Kanatchikov95a,Kanatchikov98a,Kanatchikov98b,%
Kanatchikov98c,Kanatchikov98d,Kanatchikov99a,Kanatchikov00a,%
Kanatchikov01a,Kanatchikov01b}
  and the Clifford algebra generated by \(c_a\)---to the vertical.
\end{rem}

\subsection*{Acknowledgement} 
I am very grateful to Prof.~I.V.~Kanatchikov for introducing me to the
De~Donder--Weyl theory and many stimulating discussions and comments.
Dr.~Gong Yafang also read the paper and made several useful remarks
which are gratefully acknowledged. Last but not least the anonymous
referee suggested many important improvements and corrections. The
first three paragraphs of Introduction are mainly borrowed from the
referee's report.

\newcommand{\noopsort}[1]{} \newcommand{\printfirst}[2]{#1}
  \newcommand{\singleletter}[1]{#1} \newcommand{\switchargs}[2]{#2#1}
  \newcommand{\irm}{\textup{I}} \newcommand{\iirm}{\textup{II}}
  \newcommand{\vrm}{\textup{V}}
  \providecommand{\cprime}{'}\providecommand{\arXiv}[1]{\eprint{http://arXiv.o%
rg/abs/#1}{arXiv:#1}}
\providecommand{\bysame}{\leavevmode\hbox to3em{\hrulefill}\thinspace}
\providecommand{\MR}{\relax\ifhmode\unskip\space\fi MR }
\providecommand{\MRhref}[2]{%
  \href{http://www.ams.org/mathscinet-getitem?mr=#1}{#2}
}
\providecommand{\href}[2]{#2}

\end{document}